\documentclass{sf2a-conf2012}
\usepackage{graphicx}
\usepackage{hyperref}
\usepackage[]{natbib}  
\usepackage[cyr]{aeguill}
\usepackage{epstopdf}

\def\BibTeX{{\rm B\kern-.05em{\sc i\kern-.025em b}\kern-.08em
    T\kern-.1667em\lower.7ex\hbox{E}\kern-.125emX}}
\bibpunct{(}{)}{;}{a}{}{,}  %%%%%%%%%%%%%  A&A bibliography style
\newcommand{\kms}{{\mathrm{km~s^{-1}}}}

\newcommand{\meta}{{\mathrm{[M/H]}}}
\begin{document}

\TitreGlobal{SF2A 2012}

%%-----------------------------------------------------------------
%%      the top matter
%%

\title{Gaia constraints on the Galactic thick disc}

\runningtitle{Gaia constraints on the Galactic thick disc}

\author{G. Kordopatis}\address{Institute of Astronomy, University of Cambridge, Madingley Road, Cambridge CB3 0HA, UK \\ email:~gkordo@ast.cam.ac.uk}

%% IF Author3 has the same affiliation than Author1:
%\author{C.\,E. Author3$^1$}

%% IF Author3 has its own affiliation:
%\author{C.\,E. Author3}\address{Dept. of Chess, University of Games, 35101 Las Vegas, Monaco} 

%% IF Author3 has two affiliations, the one of Author1 and a second one:
%\author{C.\,E. Author3$^{1,}$}\address{Dept. of Chess, University of Games, 35101 Las Vegas, Monaco} 

%% Keep this line, even if the page will be settled afterwards.
\setcounter{page}{237}

%%-----------------------------------------------------------------

\maketitle

%%-----------------------------------------------------------------
%%        The abstract
%% 
%%  Warning!  within the abstract:
%%  - do not use macros. 
%%  - do not use commands like: \cite, \citet, \citep ... etc.

\begin{abstract}
%Thick discs are thought to be inherent in every disc galaxy. The one of the Milky Way is mainly composed by old ( $\sim$10Gyr), intermediate metallicity ([M/H]$\sim$-0.5 dex) stars and hence can be used to probe the accretion history of our Galaxy.
The Gaia mission, with its unprecedented astrometric and photometric precision, combined with its Radial Velocity Spectrometer, will provide to the astronomical community a wealth of necessary constraints to disentangle between  the different formation scenarios  of the Galactic thick disc. The aim of this review is to present some of the recent results obtained spectroscopically concerning this Galactic structure, and highlight the open questions that still remain to be answered under the Gaia era. These concern mainly the measurement of the chemo-dynamical properties of the Milky Way at the inner and outer parts, which allow us to determine the total accreted mass from the mergers with satellite galaxies, and will give us an estimate of the strength of the radial migration phenomena to form such a structure. 
%These questions are going to be related to the ongoing and future international spectroscopic projects like RAVE, the Gaia-ESO survey and Gaia, in order to show how these missions will help to answer them.
\end{abstract}

%% Insert the keywords (to appear in the ADS indexing)
%% Keywords must be separated by a comma
\begin{keywords}
Surveys:~Gaia -- Techniques:~ spectroscopy -- Galaxy:~structure, stellar content, evolution
\end{keywords}

%%-----------------------------------------------------------------

\section{Introduction}
%%---------------------
Surveys of external  galaxies seem to suggest that thick discs are inherent structures in most (if not all) disc galaxies \citep{Yoachim08,VanderKruit11}. Although the existence of such a structure for the Milky Way has been highlighted for almost thirty years now \citep{Yoshii82,Gilmore83}, its origin is still uncertain, and many scenarios have been proposed to explain it. These can be separated into those involving internal mechanisms or those requiring external accretion or trigger in order to form the thick disc \citep[e.g.:][ and references therein]{Abadi03, Brook04, Villalobos08, Loebman11}. 

Most of the stellar spectroscopic and photometric surveys of stars in the Milky Way have shown that the Galactic thick disc is mainly composed by old stars \citep[$\sim 10$~Gyr,][]{Fuhrmann08}, of intermediate metallicity \citep[$\meta \sim-0.5$~dex, e.g.:][]{Bensby07,Kordopatis11b} and with hotter kinematics compared to the thin disc stars \citep[e.g.:][]{Casetti-Dinescu11}. In addition, the thick disc  stars inside the solar cylinder (7$<R<$9~kpc) have a ratio of $\alpha$-element abundances over iron ($[\alpha/\rm{Fe}]$) which is enhanced compared to the $[\alpha/\rm{Fe}]$ ratio of the thin disc stars \citep[e.g.:][]{Fuhrmann08,Navarro11}. This property suggests that the thick disc has been formed in relatively short timescales ($\sim 1$~Gyr), and hence the understanding of its formation offers us the possibility to decipher the merging history of our Galaxy back to redshifts of $z\sim 1.5-2$ \citep{Freeman02}. This exploration of the origins of the Milky Way is widely known as Galactic archaeology.

Most of the models and simulations which have been proposed in order to explain the formation of the thick disc manage to successfully reproduce the locally measured properties of this structure. Nevertheless, this success has been preventing us to really disentangle between the models, since the community was lacking  large observational datasets far from the solar neighbourhood, where the models differ the most.  

The advent of multi-object spectrographs (such as VLT-FLAMES), combined with the part-time dedication of large telescopes in order to make deep and statistically significant stellar surveys (e.g.:~Gaia-ESO Survey, RAVE, SEGUE) has already changed our view of the Milky Way. The simplistic approach consisting to consider that the thick disc has formed only by one mechanism is most probably out of date. The true question that one would like to answer now, is what is the relative importance of each of these processes. In that sense, the Gaia mission \citep{Perryman01} will be a goldmine, in order to extract all the necessary information, since the satellite will map the 3D positions and kinematics, and obtain estimations of the chemical compositions of several hundreds of millions of stars.

In the following section (Sect.~\ref{Sect:scenarios}) we will briefly review the most commonly cited formation mechanisms of the thick disc. Then, in Sect.~\ref{Sect:Results} we will see how these scenarios have been constrained, by enumerating the latest results that have been obtained in terms of orbits, metallicity and kinematic gradients. This section will end up with the open questions that are still remained to be answered, and show how Gaia will help to decipher the formation history of the thick disc  (Sect.~\ref{Sect:Gaia}).

\section{Formation of the thick disc: {\it in situ} versus external mechanisms }
%%-------------------------
\label{Sect:scenarios}
One of the first scenarios that has been proposed to explain the difference in kinematics and chemistry of the stellar population composing the thick disc is the co-planar accretion of all the necessary stellar content from dwarf satellites \citep{Statler88,Abadi03}.  Such a scenario is mainly defended by the hierarchical formation history of the galaxies in the $\Lambda$CDM paradigm, where the dwarf galaxies are thought to be the building blocks of larger galaxies like ours. The direct accretion scenario predicts the absence of radial or vertical chemical gradients for the thick disc, since only one population is composing this structure. In addition, the kinematics of the stars, especially at the outskirts of the Galaxy, will depend on the inclination angle of the merger as well as on the total mass of the satellite. 

Nevertheless, the numerical simulations of such accretions have shown difficulties in the preservation of the existence of the thin disc after the merger. Models involving multiple minor mergers of satellites, which would heat dynamically the pre-existent thin disc in order to form the thick disc  have hence been developed \citep{Quinn93, Villalobos08}. In that case, the stellar population composing the thick disc is in majority the one of the thin disc at the epoch of the bombardment ($z\sim1.5$) and chemical or kinematical gradients could have persisted in the thick disc only if the initial thin disc had any of those.

Similarly, in order to explain the presence of the thin disc despite the massive accretion, \cite{Jones83} and \cite{Brook04, Brook07} have proposed that the accretion consisted of a gas-rich merger, from the collapse of which the stars forming the existent thick disc will have formed. In that configuration, the stars are born {\it in situ}, but using the accreted extra-galactic gas. The orbital and chemical properties of the thick disc stars, such as radial and vertical gradients, could then be explained by the time-scale of the cloud collapse and the accretion parameters of the merger. 

Finally, the last family of scenarios require no external trigger at all. Here, the stars composing the thick disc are formed entirely in the thin disc or in the bulge and are  moved afterwards far from the Galactic plane due to internal mechanisms. 
Such a mechanism is presented for example by \cite{Bournaud09}, where it is suggested that at high redshift the turbulent primordial discs would had scattered far from the plane part of the  gas and stars, forming in that way the thick discs of external galaxies as we see them today. Like for the direct accretion or the gas-rich merger, it is predicted in such a scenario that the thick disc is chemically homogeneous. More interestingly, the inside-out evolution of the thin disc  will have as a consequence to form a thick disc with a shorter scale-length than the one of the thin disc.

Another evoked internal mechanism in order to create a thick disc is the radial migration of the stars due to resonances with the spiral structure or the bar of the Milky Way   \citep[e.g.:][]{Schoenrich09, Minchev10, Loebman11, Bird12, Minchev12}. In such scenarios, the thick disc stars inside the solar cylinder  would had come in majority from the inner radii of the Galaxy, gaining vertical velocity due to the lower Galactic potential at the outskirts, keeping a chemical signature of the interstellar medium of the regions where they have been formed. Nevertheless, the assumed dynamical evolution of the disc  implies that the information concerning the origin of the stars is blurred because of the radial migration, and hence that there is no distinctive thin and thick disc components: the disc is characterised by smooth changes in its chemo-dynamical properties. 

\section{The observational constraints at the pre-Gaia era: The unlike scenario of a single mechanism}
%---------------------------------------------------------
\label{Sect:Results}
The advent  of spectroscopic surveys  either dealing with hundreds of stars \citep[e.g.][]{Fuhrmann08, Bensby07, Kordopatis11b}, or the ones organised around big collaborations with few hundreds of thousand stars (e.g.: SEGUE, RAVE) have already brought a wealth of information in order to claim nowadays that the formation history of the Milky Way's thick disc is very likely a combination of most of the above cited scenarios.

\begin{figure}[t!]
 \centering
 \includegraphics[width=0.48\textwidth,clip]{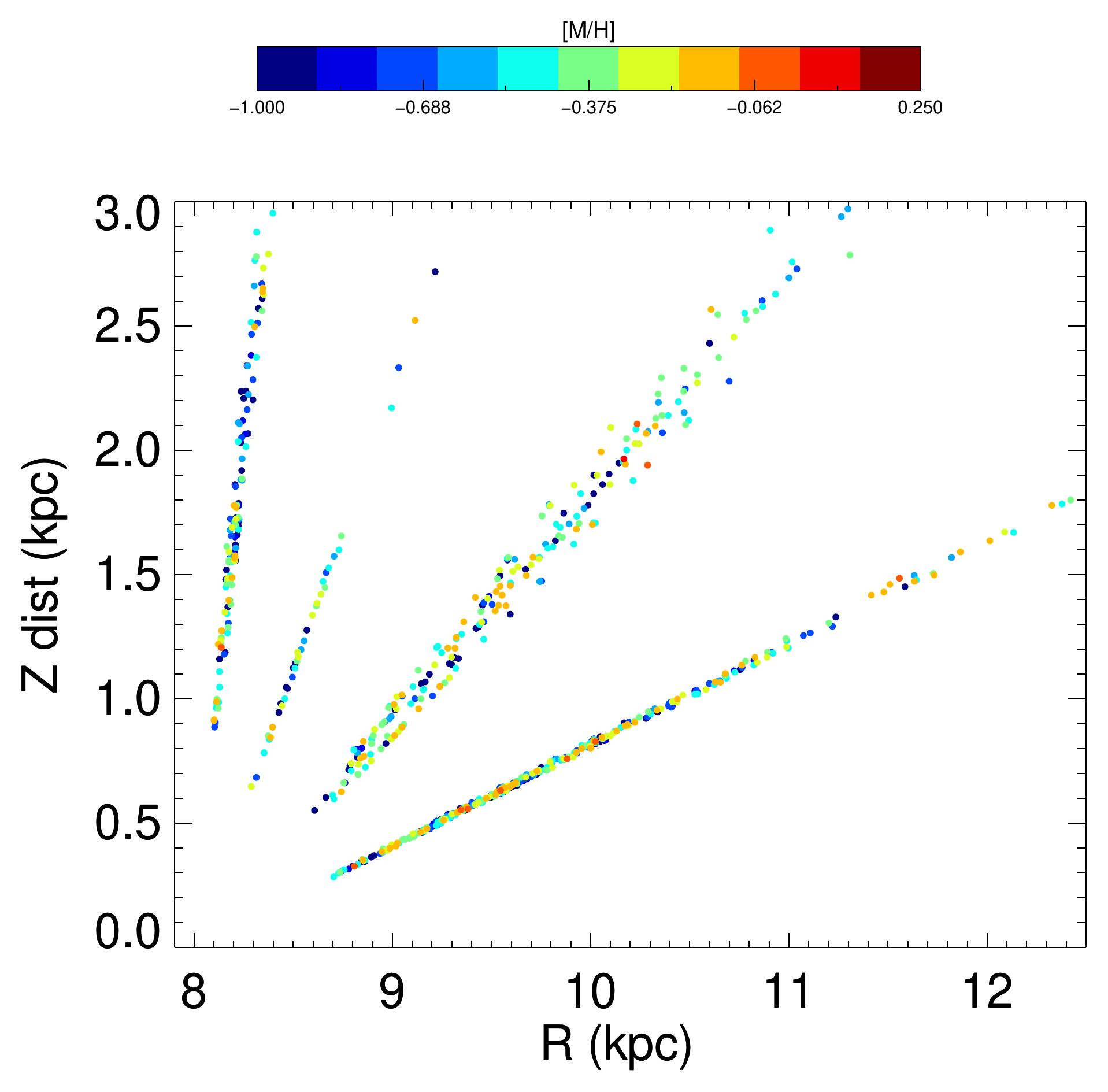}%      
 \includegraphics[width=0.48\textwidth,clip]{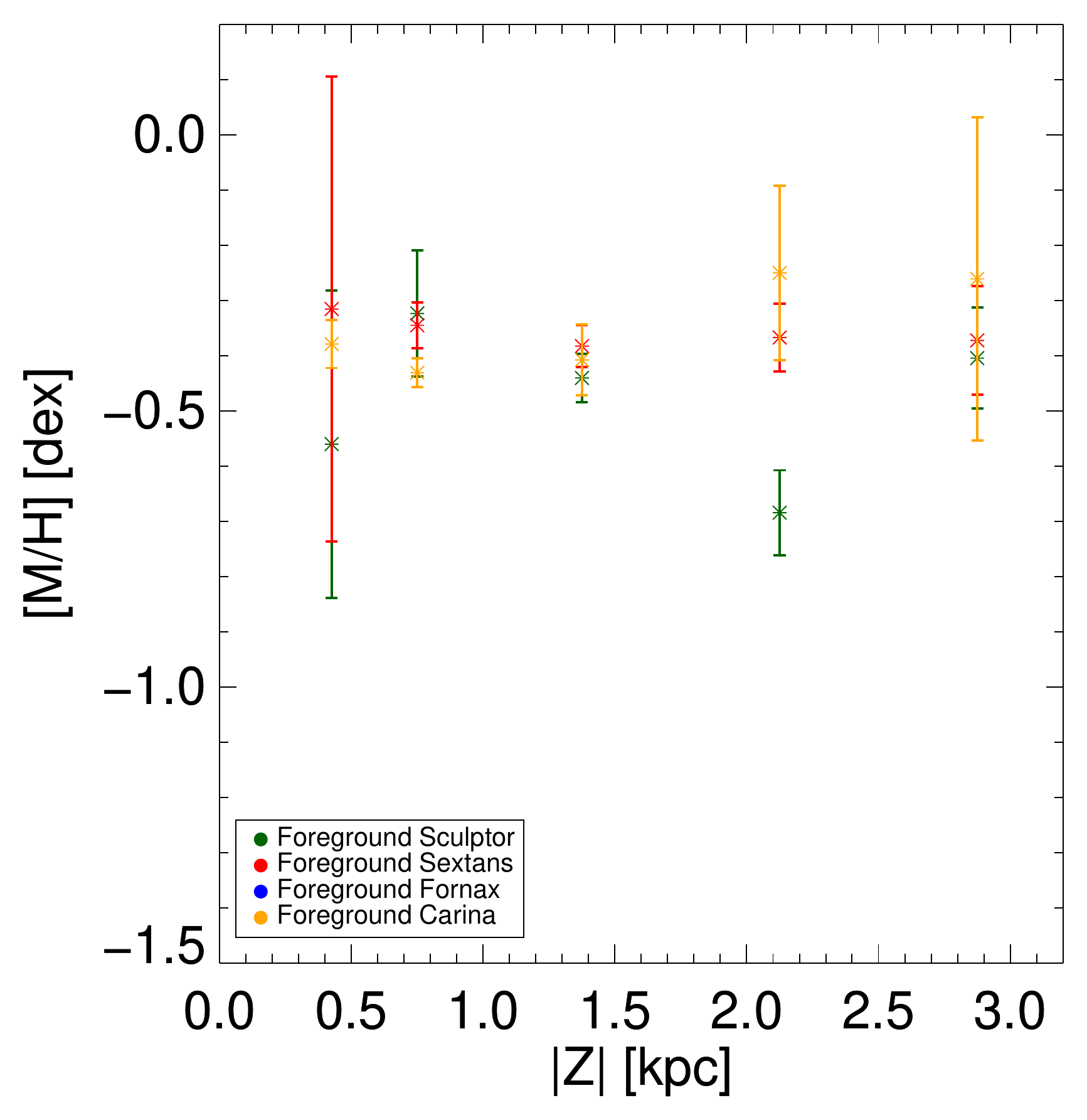}      
%% Note the ABSENCE of the extension .pdf , .eps or .ps  !
  \caption{Preliminary results of the study of $\sim$1200 foreground stars observed towards  the line-of-sights of the dwarf spheroidal galaxies of Sculptor, Fornax, Sextans and Carina (Kordopatis et al. 2012, in prep.). {\bf Left:} Metallicity of the surveyed stars at different distances above the Galactic plane and different galatocentric radii.  {\bf Right:} Measured vertical metallicity gradients  towards the line-of-sights for which enough stars were available (Sculptor in green, Sextans in red, Carina in yellow). The preliminary results seem to suggest a chemically homogeneous thick disc in the studied directions. }
  \label{Fig:DART_foreground}
\end{figure}

For instance, \cite{Cheng12a} used SEGUE data of $\sim7000$ stars to measure the radial metallicity gradients of the thin and thick discs, at different distances above the Galactic plane, up to $|Z|\sim1.5$~kpc. They have shown that the farther from the plane, the flatter the gradient, suggesting that the thick disc population is chemically homogeneous, in agreement with the scenario of \cite{Bournaud09}, the one of \cite{Abadi03} or with a radial migration mechanism (provided a strong migration rate). 
In a complementary approach, the analysis of $\sim$1200 foreground stars of several spectroscopic surveys towards the line-of-sights of the dwarf spheroidal galaxies of Sculptor, Fornax, Sextans and Carina (Kordopatis et al. 2012a, in prep) has yielded to the measurement of the vertical chemical gradient for the thick disc stars towards these four directions. The preliminary results of this study also seem to point towards a chemically homogeneous thick disc (see Fig.~\ref{Fig:DART_foreground}), in agreement with the result of \cite{Cheng12a}.

Nevertheless, a pure radial migration scenario as being the fundamental one in order to form the thick disc seem to be ruled out when analysing the eccentricity distribution function of the thick disc stars. The RAVE results \citep{Wilson11}, the SEGUE ones  \citep{Dierickx10,Lee11} as well as the independent ones of  \cite{Kordopatis11b}, obtained far from the solar neighbourhood, have shown that the thick disc stars have a peak at intermediate values ($\epsilon \sim 0.3$), with a tail going towards higher values, where merger scenarios predict that the accreted stars are found \citep{Sales09}. In addition, \cite{Lee11} have shown that  there is a strong correlation between the metallicity and the eccentricity of the stars, a characteristic which does not seem to be existent for the thin disc stars, hence suggesting that radial migration played a less important role for the thick disc than for the thin one. 

Similarly, the strong correlation between the rotational velocity and the metallicity measured by \cite{Kordopatis11b, Lee11, Schlesinger11} ($\partial V_\phi / \partial \meta \sim 45~\kms$~dex$^{-1}$) 
  seem to decrease again the relative importance of the radial migration in the formation of the thick disc\footnote{We note, however, that recently \cite{Curir12} suceeded to simulate such a strong correlation using radial migration in a barred disk Galaxy}. This correlation, combined with the measurement of the vertical metallicity gradient of more than $-0.1$~dex~kpc$^{-1}$ \citep{Kordopatis11b, Ruchti11}, as well as the similarity of the [$\alpha$/Fe] ratio of the  metal-poor thick disc stars with the [$\alpha$/Fe] ratio of the halo stars \citep{Ruchti10}, also rules-out the pure accretion scenario. 
%  The absence of scatter in the $\alpha-$element abundances of 234 metal-poor stars of the thick disc, found by \cite{Ruchti10}, as well as their agreement with the [$\alpha$/Fe] ratio of the halo stars also 

With such a reasoning, and using a sample of $\sim 12\times 10^3$ G-dwarfs, \cite{Liu12} have suggested that the thick disc could be decomposed into a radial migrated population being $\alpha-$old and having circular orbits, and another one formed from extra-galactic material, composed with the most metal-poor stars ([M/H]$<-0.6$~dex) on eccentric orbits. 
Worth mentioning are also the studies of \cite{Bensby11} and \cite{Cheng12b} who found no $\alpha-$enhanced stars at the outskirts of the galaxy, at the distances above the plane where the thick disc should be the dominant population. Whereas both studies rely on relatively poor statistics at these radii, both of them have suggested a smaller scale-length for the $\alpha-$enhanced thick disc. In particular, \cite{Cheng12b} proposed as an explanation that the inner part of the thick disc could be consistent with a scenario in which the thick disc forms during an early gas-rich accretion phase. Furthermore, the stars far from the plane in the outer disc could have reached their current locations through heating by minor mergers. The precise measurement of the thick disc properties at these radii could hence impose important constraints on the strength of radial migration in the thin disc.

%The questions that someone would like to answer, is what is the relation between the thick disc and the other galactic components of the MW, like the thin disc, the bulge or the halo?  
%In particular, we would like to know how much different is the Thick Disc from the thin disc \citep[][]{Bovy11,Liu12} or the bulge \citep{Bensby??}? 

\section{The Gaia legacy: determining the relative importance of each formation scenarios}
%-----------------------------------------------
\label{Sect:Gaia}

The Gaia satellite is planned to be launched in September 2013. 
During its five year mission, the survey will aim for completeness to $V \sim 20-25$~mag,  depending on the colour of the object, with astrometric accuracies of about 20~$\mu$as at $V\sim$15~mag. Within a radius of 10~kpc from the Sun, the 3D motions and positions of several hundreds of millions of stars will be known with accuracies better than 10\%.  %In the process, it will map the stellar motion and provide the detailed physical properties of each star observed: characterizing their luminosity, temperature, gravity and elemental composition.
The onboard Blue (BP) and Red (RP)  spectro-photometers will deliver broadband photometry for roughly a billion stars, up to the  20th magnitude.  For the brightest targets ($V\leq 17$~mag, $\sim 350 \times 10^6$~stars), these data will be complemented by the intermediate resolution spectra ($R\sim11~500$) gathered at the near-infrared, around the ionised Calcium triplet ($\lambda \lambda~847 - 874$~nm) from the Radial Velocity Spectrometer (RVS).
The analysis of these spectra, will allow to obtain the effective temperature ($\rm T_{\rm eff}$), surface gravity ($\log g$), overall metallicity and $\alpha-$abundances for the stars stars brighter than $V\sim14$~mag. In particular, \cite{Kordopatis11a} have estimated that accuracies of 100~K, 0.17~dex and 0.1~dex for the $\rm T_{\rm eff}$, $\log g$ and [M/H], respectively, could be achieved for typical thick disc stars at a signal-to-noise ratio of 50~pixel$^{-1}$ using their automated stellar spectra parameterisation pipeline.  

The first astrometry data release is likely to be in 2016, with spectrophotometry and stellar parameters to follow later, and 2021 for the final catalogue.
This wealth of data, will  of course shed some light on the relative contribution of each of the previously cited mechanisms, revealing the relation between the thick disc and the other Galactic structures (thin disc, bulge and halo). In addition, the volume and the magnitudes that are going to be surveyed by Gaia, will minimise the effects of specific spatial and stellar mass sampling at the solar cylinder from which suffered previous surveys, and which according to \cite{Bovy12}, can bias the interpretation of the distinctive nature of the thick disc. More precisely, the constraints that Gaia will bring on the formation scenarios of the thick disc  are the followings:  
\begin{itemize}
\item The observation of the outskirts of the galaxy ($R>8$~kpc) will allow to obtain estimations of the scale-length of the thick disc, both in terms of star counts, but also using a Jeans analysis. The measurement of a flaring for the thick disc will be possible, which will give valuable information about the total accreted mass of the Milky Way \citep{Qu11}. In addition, the strength of the radial migration processes will be characterised robustly, by analysing the chemical homogeneity at the outer parts of the thick disc with a statistically significant stellar sample.  \\

\item  The comparison of the star counts towards the northern and the southern  Galactic cap, the search of kinematic asymmetries compared to the Galactic plane, as well as the radial gradients at different heights above the plane, will allow to detect in the thick disc the possible remnants  of satellite accretions, either as stellar streams or over-densities in the phase-space. In addition, the evolution of the ellipticity of the orbits, or the variations in the correlations between the circular orbital velocity and the metallicity at different radii, will determine if there is complete mixing in the thick disc stars \citep{Loebman11}, and perhaps constrain the influence of the Galactic bar in the radial migration of stars. \\

\item 
The observation the thick disc  towards the Galactic bulge will be limited by the high extinction of these regions, as well as the crowding limit of the RVS (20~000 stars~deg$^{-2}$). Nevertheless, the chemo-dynamical characterisation of the brightest thick disc giants in that direction will reveal if there is a relation between the bulge stars and the thick disc ones, which will be for the first time a direct proof of radial migration as one of the mechanisms forming the thick disc structure.  

%By observing the chemo-dynamical properties of the brightest thick disc giants in direction of the Galactic bulge (because of the high extinction and the crowding limits of the RVS), we will be able to find if there is a relation between the bulge stars and the thick disc ones, which will be for the first time a direct proof of radial migration as one of the mechanisms forming the thick disc structure. 
\end{itemize}

Finally, it should be noticed that the results of Gaia will be complemented by the already on-going Gaia-ESO public spectroscopic survey \citep{Gilmore12}. This project employs the VLT-FLAMES instrument for high quality spectroscopy of some 100~000 Milky Way stars. Targets  brighter than $V\sim18$~mag will be observed  with GIRAFFE at a resolution of  $R\sim$20~000 whereas some 10~000 stars brighter than $V\sim16$ will have UVES spectra at a resolution of  $R\sim45~000$.  At the end of this survey, it is estimated that roughly 20~000 stars belonging to the thick disc will be observed, complementing the Gaia mission with more precise stellar radial velocities and chemical abundances at the fainter magnitudes.

\section{Conclusions}
\label{Sect:Conclusions}
The understanding of the formation of the thick disc is directly linked with our understanding of the hierarchical formation of the galaxies in the Universe. This review presented the most commonly evoked scenarios to form the Galactic thick disc, and highlighted which are the most recent findings from the large spectroscopic surveys such as RAVE or SEGUE.  The future Gaia mission will complement the advent of the future data releases from RAVE (Kordopatis et al. 2012b, to be submitted), SEGUE, APOGEE and Gaia-ESO survey, by mapping chemo-dynamically the regions of the thick disc in which the formation scenarios differ the most.  The combination of all these projects will hence determine the relative importance of internal and external mechanisms for the origin and the evolution of the thick disc.

% Optional acknowledgements
% -------------------------
\begin{acknowledgements}
G.K. would like to thank the SOC for the invitation in order to give this presentation and the AS-Gaia consortium for financial support. G.~Gilmore and P.~de~Laverny are warmly thanked  for the careful reading of this proceeding.
\end{acknowledgements}

\bibliographystyle{aa}  % A&A bibliography style file (aa.bst)
\bibliography{Kordopatis} % your references in file: Yourfile.bib

\end{document}